\documentstyle[12pt]{article}

\begin{document}

\centerline{\large \bf FERMION MASSES AND MIXINGS IN SO(10) GUTS }  

\vskip 10truemm
\centerline {\bf Borut Bajc$^1$}  
\vskip 5truemm
\centerline{${}^1$ {J.Stefan Institute, 1001 Ljubljana, Slovenia}}

{\renewcommand{\thefootnote}{}
\footnote {\hskip -5mm Based on talks given at the 2005 Balkan Workshop, 
Vrnja\v cka Banja (Serbia), 19-23 May 2005 (to be published in the 
Proceedings) and at PASCOS 2005, Gyeongju (Korea), 30 May-4 June 2005 
(published in AIP Conf.Proc.805:326-329,2006)}
}


\begin{quote}
\small{\bf Abstract.} {\it I will present a simple, economic 
and predictive model of Yukawa 
structures in the context of a renormalizable SO(10) grandunification. 
The righthanded neutrino mass is generated radiatively. The 
fermions have Yukawa couplings with one $10$ and one $120$ dimensional 
Higgses. In the approxiamate two generation scenario the model 
predicts degenerate neutrinos, small quark mixing and $b-\tau$ 
Yukawa unification.}
\end{quote}

\section{Introduction}

In the following talk I will study some simple example 
of flavour physics in the context of SO(10) grandunificastion 
alone, i.e. assuming no flavour symmetry form the beginning. 
Clearly such a model would be highly undetermined and arbitrary, 
if we considered all the operators allowed by SO(10). The reason 
for this is that at least the first generation of fermion masses 
could be influenced (if not dominated) by the operators suppressed 
only by $M_{GUT}/M_{Planck}\approx 10^{-3}$. Thus we are forced 
to limit ourselves to renormalizable terms only, hoping that for 
some strange and unknown reason Planck physics does not generate 
this $1/M_{Planck}$ suppressed operators. That this may not be so crazy 
is enough to remember that in any supersymmetric grandunification 
a term 

\begin{equation}
W=\frac{c}{M_{Planck}}QQQL\;\;\;(\subset 16_F^4)
\end{equation}

\noindent
would be generically allowed. Constraints from $d=5$ proton decay 
rules out all such theories except those with a negligible coefficient 

\begin{equation}
c\leq 10^{-7}
\end{equation}

In a similar spirit we assume that all nonrenormalizable operators 
are negligible. 

The model \cite{Bajc:2005aq,Bajc:2004hr} 
I will present is a counterexample to the following 
often claimed statements:

\begin{itemize}
\item
   we need the $126_H$ Higgs representation (or the $16_H$ plus the 
   $1/M_{Planck}$ suppressed operators), to give 
   mass to the right-handed neutrino 
\item
   $b-\tau$ unification of Yukawa couplings follows only in 
   models with $10_H$ domination
\item
   small mixing angles in the quark sector and at the same time 
   large mixing angles in the leptonic sector seem difficult to achieve 
   and are considered as problems or at least mysteries to be explained
\item
   prediction of hierarchical neutrinos 
\end{itemize}

\section{The model}

The most general renormalizable Yukawa terms in SO(10) can be 
schematically written as 

\begin{equation}
\label{ly}
{\cal L}_Y=16_F^T\left(10_HY_{10}+120_HY_{120}+126_HY_{126}\right)16_F\;,
\end{equation}

\noindent
where from SO(10) algebra alone one can determine the symmetricity 
and antisymmetricity of the $3\times 3$ Yukawa matrices:

\begin{equation}
\label{y}
Y_{10,126}^T=+Y_{10,126}\;\;,\;\;Y_{120}^T=-Y_{120}\;.
\end{equation}

The most general case is not restrictive, so one tries simple models, 
minimal subcases, which could be potentially realistic.

What about the Higgs sector? There are two types of Higgs representations 
that break the rank of SO(10). The first one is the already mentioned 
$126_H$. We want a counterexample to the first item in the introduction, 
so we will avoid this representation and choose instead $16_H$. The 
vev of this representation alone breaks SO(10) to SU(5), so it could 
give in principle a large SU(5) invariant mass to the righthanded 
neutrino. The problem is that (\ref{ly}) does not contain any $16_H$, 
i.e. this Higgs representation can be coupled to the fermionic $16_F$ 
only through a nonrenormalizable operator. Since we assumed these 
to be absent, at tree level we get the righthanded neutrino massless, 
$M_{\nu_R}=0$. We can however avoid failure by recalling the old Witten's 
idea \cite{Witten:1979nr}: at two loop level there are diagrams 
that generate an effective operator of the form 

\begin{equation}
\label{effop}
\frac{16_F^216_H^2}{M}\;.
\end{equation}

Witten took as an example the diagram with gauge bosons exchange and  
$10_H$ coupled to fermions. The mass scale $M$ in (\ref{effop}) 
is presumably the heaviest mass in the loop, something like $M_{GUT}$. 
Since there is a two loop suppression, the righthanded neutrino mass 
matrix become

\begin{equation}
\label{mnr}
M_{\nu_R}\approx \left(\frac{\alpha}{\pi}\right)^2
\frac{M_R^2}{M_{GUT}}Y_{10}\;,
\end{equation}

\noindent
where $M_R=\langle 16_H\rangle$ is the scale of the SU(2)$_R$ breaking. 

When is this approximate formula valid? 

In a nonsupersymmetric model this will hardly work. The reason is 
that there is no one-step gauge unification in such theories, so that an 
intermediate scale is needed. Usually this scale is exactly $M_R$, 
and typically it lies few orders of magnitude below $M_{GUT}$. This 
would further suppress the already two loop suppressed righthanded 
neutrino mass, predicting an unacceptably large light neutrino masses, 
well above the approximate limit of $1$ eV. This of course assuming 
that the Dirac neutrino Yukawa matrix has not all elements small, 
i.e. that there are similarity relations between the up quark 
Dirac neutrino mass matrices, as is usually the case in SO(10). 
The only possible exception is that $M_R$ is still high, but there 
are some light multiplets at an intermediate scale that make the 
running of the gauge coupling modify in the appropriate way. We 
will not consider this case anymore. 

The supersymmetric version cannot work. The righthanded neutrino mass 
term comes from a superpotential, and it is well known that in 
supersymmetry the superpotential does not get renormalized. So if 
such a term was not present at tree order, as we assumed, it will 
not appear at any order of perturbation theory. In other words, the 
righthanded neutrino mass must be proportional to the supersymmetry 
breaking scale, i.e. to some positive power of the ratio $m_{SUSY}/M_{GUT}$. 
In low energy supersymmetry breaking like in MSSM this is obviously 
far too small. 

Although at this point the situation seems hopeless, the above 
examples give a hint of the direction to follow. What we need is 
a one step unification together with a large susy breaking scale. 
Such a model has been proposed last year, and it is nothing else 
than split supersymmetry \cite{Arkani-Hamed:2004fb}. 
In these scenarios the running 
of the gauge couplings gets improved by the presence at low energy of 
new fermions like gauginos and higgsinos. The absence at the low energy 
of the sfermions is not important since they come along in SU(5) 
multiplets and thus do not contribute at the one loop level to the 
determination of the unification scale. In the original proposal 
these supersymmetric partners of the SM fermions could have any mass 
between $M_W$ and $M_{GUT}$. If we want to generate 
the righthanded neutrino masses with the radiative mechanism, we 
need them very massive, i.e. at $m_{SUSY}\approx M_{GUT}$ 
\cite{Bajc:2004hr}, an interesting and important result. 
Although such a large susy breaking is ruled out, one can still 
play with some degree of fine-tuning. Even this would not easily 
help the nonsupersymmetric version, due to the wrong prediction 
of $b-\tau$ unification (see below).

The now missing $126_H$ representation had also another function, i.e. 
to correct the Yukawa matrices. In fact the standard model Higgs 
is in general a linear combination of SU(2)$_L$ doublets coming from 
different representations, and the two most important contributions 
were certainly those coming from $10_H$ and $126_H$ \cite{Babu:1992ia}. 
This option is now gone, and one needs to add another Higgs representation 
to the Yukawa sector, on top of $10_H$. From (\ref{ly}) we see as the 
first possibility to add a $120_H$. This is what we will do in detail 
in the next chaper.

\section{$10_H+120_H$}

The matrix $Y_{120}$ is very restrictive, since it contains 
only 3 complex parameters, being antisymmetric (\ref{y}). 
Partially one gain some new parameters with respect to the case 
$10_H+126_H$ \cite{Babu:1992ia} due to some new doublet vevs. In fact 
we have 6 unknown SU(2)$_L$ breaking vevs that contribute to the 
light fermion masses:

\begin{eqnarray}
\langle(2,2,1)_{10}\rangle&=&\pmatrix{
  v_{10}^u
& 0
\cr
  0
& v_{10}^d
\cr}\;,\\
\langle(2,2,1)_{120}\rangle&=&\pmatrix{
  v_{120}^u
& 0
\cr
  0
& v_{120}^d
\cr}\;,\\
\langle(2,2,15)_{120}\rangle&=&\pmatrix{
  w_{120}^u
& 0
\cr
  0
& w_{120}^d
\cr}\;.
\end{eqnarray}

Can we fit the data with $10_H$ and $120_H$ only? We will see 
that in the two generation case the answer is yes. But 
let us start more generally. The mass matrices that follow from 
(\ref{ly}) without $126_H$ after symmetry breaking are 

\begin{eqnarray}
M_D&=&v_{10}^dY_{10}+\left(v_{120}^d+w_{120}^d\right)Y_{120}\;,\\
M_U&=&v_{10}^uY_{10}+\left(v_{120}^u+w_{120}^u\right)Y_{120}\;,\\
M_E&=&v_{10}^dY_{10}+\left(v_{120}^d-3w_{120}^d\right)Y_{120}\;,\\
M_{\nu_D}&=&v_{10}^uY_{10}+\left(v_{120}^u-3w_{120}^u\right)Y_{120}\;,\\
M_{\nu_R}&=&v_RY_{10}\;,\\
M_N&=&-M_{\nu_D}^TM_{\nu_R}^{-1}M_{\nu_D}\;.
\end{eqnarray}

In the above equation $v_R$ is just the coefficient of the 
two loop contribution (\ref{mnr}). We have used the type I seesaw 
formula, because the type II seesaw contribution $M_{\nu_L}$ is 
two loop suppressed similarly as $M_{\nu_R}$, while the type I, 
being inversely proportional to $M_{\nu_R}$ gets at the same time 
enhanced. One could say that the type II over type I contribution 
is proportional to $(\alpha /\pi)^4$, so I will neglect the type II 
contribution from now on. 

The above formualae can be rewritten in a useful way:

\begin{eqnarray}
M_D&=&M_0+M_2\;,\\
M_U&=&c_0M_0+c_2M_2\;,\\
M_E&=&M_0+c_3M_2\;,\\
M_{\nu_D}&=&c_0M_0+c_4M_2\;,\\
M_{\nu_R}&=&c_RM_0\;,
\end{eqnarray}

\noindent
where $M_0$ is symmetric and $M_2$ is antisymmetric. 

The full analysis is hard and time consuming. We will show first 
what happens in the approximate world of two generations. This in order 
to get some analytic results. Consider then the $2^{nd}$ 
and $3^{rd}$ generations only. In all generality we can choose a 
basis, in which $M_0$ is diagonal with positive real eigenvalues:

\begin{eqnarray}
M_0=\pmatrix{
  a
& 0
\cr
  0
& b
\cr}\;,\;
M_2=\pmatrix{
  0
& -i\alpha
\cr
  i\alpha
& 0
\cr}\;.
\end{eqnarray}

In general $a$, $b$ and $c_{0,R}$ can be taken real ($a,b>0$), while 
$\alpha$, $c_{2,3,4}$ are complex. 


\subsection{$b-\tau$ unification}


This prediction is relatively easy to obtain. In practice it 
follows because both $M_D$ and $M_E$ has the same diagonal elements, 
and because the second generation of fermions is much lighter than 
the third one. 

Since the above matrices are not Hermitian in general, one better 
calculates 

\begin{eqnarray}
\label{memn}
M_DM_D^\dagger &=&\pmatrix{
  a^2+|\alpha |^2
& -i(a\alpha^*+b\alpha)
\cr
  i(a\alpha+b\alpha^*)
& b^2+|\alpha |^2
\cr}\;.
\end{eqnarray}

From the trace and the determinant of the above matrix 
one finds for the down quark masses

\begin{eqnarray}
\label{traced}
a^2+b^2+2|\alpha |^2&=&m_b^2+m_s^2\;,\\
\label{detd}
|ab-\alpha^2|^2&=&m_b^2m_s^2\;.
\end{eqnarray}

Similar relations can be obtained for the charged leptons, 
just change $\alpha\to c_3\alpha$, $m_b\to m_\tau$, $m_s\to m_\mu$. 

Writing $\alpha=|\alpha|e^{i\phi}$ and taking into account that 
$\cos{(2\phi)}\le 1$, one finds that the sum $a+b$ is constrained:

\begin{equation}
\label{abd}
m_b-m_s\le a+b\le m_b+m_s\;.
\end{equation}

For the charged lepton sector one gets a similar bound,

\begin{equation}
\label{abe}
m_\tau-m_\mu\le a+b\le m_\tau+m_\mu\;.
\end{equation}

Clearly, in order to satisfy both requirements (\ref{abd}) and 
(\ref{abe}) the intervals $[m_b-m_s,m_b+m_s]$ and $[m_\tau-m_\mu,
m_\tau+m_\mu]$ must have at least some region in common. Thus the 
bottom and tau masses can differ by less than 

\begin{equation}
|m_b-m_\tau | \le m_s+m_\mu\;,
\end{equation}

\noindent
i.e. we get $b-\tau$ unification with a quite good precision.

It has to be stressed, that this $b-\tau$ unification has a very 
unusual origin: it is not due to the dominance of the Higgs in 
the 10-dimensional representation as it is usually the case, but it is 
a direct consequence of the smallness of the second generation 
charged fermion masses $m_s$ and $m_\mu$.

\subsection{Degenerate neutrinos}

Let's now prove, that a large leptonic mixing angle forces 
the neutrinos to be nearly degenerate. The main point is the 
particular form of $M_2$, which is just the second Pauli matrix. 
This makes the calculation of $M_N$ particularly easy, with 
the surprising result that the neutrino mass matrix is diagonal:

\begin{equation}
M_N=c_NM_0\;\;,\;\;c_N=\frac{c_4^2\alpha^2-c_0^2ab}{abc_R}\;.
\end{equation}

This was obtained without any approximation, except for 
working in the two generation case.

Similarly as before we calculate 

\begin{eqnarray}
\label{me}
M_EM_E^\dagger &=&\pmatrix{
  a^2+|c_3\alpha |^2
& -i(ac_3^*\alpha^*+bc_3\alpha)
\cr
  i(ac_3\alpha+bc_3^*\alpha^*)
& b^2+|c_3\alpha |^2
\cr}\;,\\
\label{mn}
M_NM_N^\dagger &=&|c_N|^2\pmatrix{
  a^2
& 0
\cr
  0
& b^2
\cr}\;.
\end{eqnarray}

From the trace and the determinants of the above matrices 
one finds for the lepton masses ($\phi_3$ is the phase of 
$c_3\alpha$)

\begin{eqnarray}
\label{tracee}
a^2+b^2+2|c_3\alpha |^2&=&m_\tau^2+m_\mu^2\;,\\
\label{dete}
a^2b^2+|c_3\alpha|^4-2ab|c_3\alpha|^2\cos{(2\phi_3)}
&=&m_\tau^2m_\mu^2\;,\\
\label{m1}
|c_N|^2a^2&=&m_2^2\;,\\
\label{m2}
|c_N|^2b^2&=&m_3^2\;.
\end{eqnarray}

The relative angle between the matrices in (\ref{memn}) is nothing 
else than the neutrino atmospheric mixing angle, and it is easily 
calculated from the invariant 

\begin{eqnarray}
\label{trmemn}
Tr\left(M_EM_E^\dagger M_NM_N^\dagger \right)&=&
m_\tau^2m_3^2+m_\mu^2m_2^2\\
&-&\sin^2{\theta_A}
(m_\tau^2-m_\mu^2)(m_3^2-m_2^2)\nonumber
\end{eqnarray}

Using (\ref{me})-(\ref{mn}) one gets 

\begin{equation}
\label{atm}
\frac{b^2-a^2}{m_\tau^2-m_\mu^2}=\cos{(2\theta_A)}\;.
\end{equation}

One can calculate $\cos{(2\phi_3)}$ from (\ref{dete}) and 
similarly as in the previous section require that its square 
is smaller than $1$. This brings to the consistency relation

\begin{equation}
\label{c3a}
\left(m_\tau-m_\mu\right)^2\frac{\sin^2{(2\theta_A)}}{4}
\le |c_3\alpha|^2\le
\left(m_\tau+m_\mu\right)^2\frac{\sin^2{(2\theta_A)}}{4}\;.
\end{equation}

Putting all together 

\begin{equation}
\frac{m_3^2-m_2^2}{m_3^2+m_2^2}=\frac{(m_\tau^2-m_\mu^2)\cos{(2\theta_A)}}
{(m_\tau^2+m_\mu^2)-(m_\tau+\xi m_\mu)^2\sin^2{(2\theta_A)}/2}\;,
\end{equation}

\noindent
where $\xi^2\le 1$ parametrizes the value of $|c_3\alpha|$ in the 
range given by (\ref{c3a}).

This shows that for a nearly maximal atmospheric mixing angle the 
neutrinos tend to be degenerate. Of course one needs a better numerical 
check \cite{Bajc:2005zf}, but the final word can be given only after 
the three generation analysis. Irrespectively of the numerical fit, 
it is however interesting that in this model a connection exists between 
the large atmospheric mixing angle and the neutrino degeneracy. 

\subsection{Small quark mixing angles}

Another interesting connection exists between the large atmospheric 
mixing angle and the small quark mixing angle. This relation is a bit 
different from the ones we derived before. In fact, so far, although 
we kept all the masses nonzero, basically very similar results are 
obtained in the limit of massless second generation charged fermions. 
In such a limit however the quark mixing angle becomes exactly zero 
\cite{Bajc:2005aq}. Although this is a good approximation in the 
leading order, it would be interesting to get the main nonzero 
contribution, as it was done in \cite{Bajc:2005zf}

To find the quark mixing angle $\theta_{cb}$ 
one can use the quark sector analogue of (\ref{trmemn}):

\begin{eqnarray}
Tr\left(M_UM_U^\dagger M_DM_D^\dagger \right)&=&
m_t^2m_b^2+m_c^2m_s^2\\
&-&\sin^2{\theta_{cb}}(m_t^2-m_c^2)(m_b^2-m_s^2)\nonumber\;.
\end{eqnarray}

A tedious, but straightforward calculation gives at leading order 

\begin{equation}
\sin{\theta_{cb}}=\xi_d\frac{m_s}{m_b}\cos{(2\theta_A)}\;,
\end{equation}

\noindent
where $\xi_d^2\le 1$ parametrizes $a+b$ in the range (\ref{abd}). 
The corrections to this formula are of higher powers of the small 
parameters $m_c/m_t$, $m_s/m_b$ and $\cos{(2\theta_A)}$. 

We found thus a relation between the large atmospheric 
mixing angle and the small quark mixing angle, together with the 
further suppression due to the small ratio of second generation masses 
to third generation masses. This is interesting per se, irrespectively 
of whether it fits the data numerically or not. In fact, the 
quark mixing angle comes out to be numerically too small. But, as 
before, further numerical analysis of the three generation case 
are needed.

\section{Conclusions}

There is by now a well defined minimal renormalizable 
supersymmetric SO(10) model \cite{Aulakh:2003kg}, 
in which the matter $16_F$ couple to the Higgs $10_H$ and $126_H$. 
It has been proved \cite{Bertolini:2005qb} 
that its Yukawa sector is consistent with 
all experimental data on light fermion masses and mixings. 
In the case of a type II seesaw dominance it predicts 
a relatively large $|U_{e3}|\ge 0.1$ \cite{Goh:2003hf} and 
gives an interesting correlation between large atmospheric 
neutrino mixing angle and $b-\tau$ unification \cite{Bajc:2002iw}.
Recent attempts to properly account the constraints from the 
known Higgs sector however indicate some inconsistencies with the 
previous solutions \cite{Bajc:2005qe}. More work is needed.

On top of that it could happen, that supersymmetry is broken at a very 
large scale, but that some fermionic partners are nevertheless 
close to the electroweak scale. In this case I offered a simple 
alternative model, in which the righthanded neutrino gets its 
mass through a radiative mechanism. The simplest Yukawa structure 
has been shown to be consistent with the second and third generations, 
predicting almost degenerate neutrinos, a correlation between 
large atmospheric mixing angle and small quark mixing angle, 
and the equality of $b$ and $\tau$ Yukawa couplings.

\vskip 5truemm

\noindent
{\bf Note added}

Yesterday the paper \cite{Aulakh:2006vi} appeared, in which some 
of the results of this paper were obtained independently. I thank 
Charan Aulakh for information of his work prior to publication 
and for correspondence.

\vskip 5truemm

\noindent
{\bf Acknowledgments}

It is a pleasure to thank the organizers of the Balkan Workshop 2005 
and PASCOS 2005. The work described above has been done 
mainly in collaboration with Goran Senjanovi\' c, although 
Alejandra Melfo and Francesco Vissani also contributed to some results. 
I thank Miha Nemev\v sek for discussion and for an earlier version of the 
proof of approximate $b-\tau$ unification. This work is supported 
by the Ministry of Education, Science and Sport of the 
Republic of Slovenia. 

{}

\end{document}